\documentstyle[amssymb,aps,pra]{revtex}

\begin{document}

\twocolumn[\hsize\textwidth\columnwidth\hsize\csname
@twocolumnfalse\endcsname
\draft
\title{Spatial fragmentation of a Bose-Einstein condensate in a double-well
potential}
\author{R. W. Spekkens and J. E. Sipe}
\address{Department of Physics, University of Toronto, 60 St. George Street, Toronto,\\
Ontario, Canada M5S 1A7}
\date{Oct. 29, 1998}
\maketitle

\widetext
\begin{abstract}
We present a theoretical study of the ground state of a Bose-Einstein
condensate with repulsive inter-particle interactions in a double-well
potential, using a restricted variational principle. Within such an
approach, there is a transition from a single condensate to a fragmented
condensate as the strength of the central barrier of the potential is
increased. We determine the nature of this transition through approximate
analytic as well as numerical solutions of our model, in the regime where
the inter-particle interactions can be treated perturbatively. The degree of
fragmentation of the condensate is characterized by the degrees of
first-order and second-order spatial coherence across the barrier.

\end{abstract}

\vfill
\narrowtext

\vskip2pc]

\section{Introduction\strut}

The recent experimental demonstration of interference phenomena in
Bose-condensed atomic gases \cite{Andrews et al.}, motivates a study of the
spatial coherence of a condensate in a double-well potential. In particular,
we are interested in the loss of spatial coherence that can occur at zero
temperature due to {\it fragmentation} of the condensate. A fragmented
condensate is one for which there is a macroscopic occupation of two or more
orthogonal single-particle wavefunctions. If the occupied single-particle
wavefunctions are spatially well separated, coherence over the spatial
extent of the entire system will be lost, persisting only over the spatial
extent of each fragment.

As Nozi\`{e}res \cite{Nozieres} has pointed out, for repulsive
inter-particle interactions it is the exchange energy that typically
prevents fragmentation into a number of degenerate (or nearly degenerate)
single-particle wavefunctions. However, this argument is inapplicable for
bosons in an external potential with several local minima, since
single-particle wavefunctions that are localized about these minima may have
very little overlap with one another, thereby leading to a very small
exchange energy. Moreover, since the self-interaction energy in such a
fragmented condensate is smaller than that of a single condensate, it is
possible for the total interaction energy to be smaller as well. Although
every particle in the fragmented condensate will pay a price in kinetic
energy to occupy localized wavefunctions, the overall energy may still be
less than that of a single condensate. Indeed, it can be shown that in the
limit of a symmetric double-well potential with an infinitely strong central
barrier, one can always find a fragmented state that has a total energy
lower than any single condensate \cite{lanl}.

Thus we have the following situation in a double-well potential: in the
absence of any central barrier the ground state is well approximated by a
single condensate, while in the presence of an infinitely strong barrier it
is well approximated by a fragmented condensate. It is clear therefore that
there must be a transition between these two extremes as one increases the
strength of the barrier. The first goal of this paper is to propose a
theoretical model for describing this transition. Specifically, \strut we
argue for an approximation of the fully interacting ground state that is
more general than a Fock state, and that can be said to describe `partial
fragmentation' of the condensate. The equations that such a state must
satisfy are derived within a variational approach. The second goal of the
paper is to solve these equations in a regime where the inter-particle
interactions can be treated perturbatively. Numerical solutions of the
equations and analytic approximations to these solutions are obtained within
this regime. It should be noted that this limit is inappropriate for the
description of the MIT condensate interference experiment \cite{Andrews et
al.}, and consequently, our results do not specify the nature of the
transition for this experimental set-up. Nonetheless, we expect the generic
features of the transition to persist in the experimentally relevant regime.

We pause to consider previous treatments of this topic and their relation to
this work. R\"{o}hrl {\it et al.}\cite{Rohrl} have provided a model of the
MIT condensate interference experiment; however, it is a mean field analysis
and therefore cannot describe fragmentation. Fragmentation in the case of
attractive inter-particle interactions has been considered by Wilkin, Gunn
and Smith \cite{Gunn}, but this effect is qualitatively different from that
of the repulsive case. Finally, Milburn {\it et al. }\cite{Milburn} have
considered the energy eigenstates of a Bose-Einstein condensate in a
double-well potential, and predict fragmentation when the inter-particle
interactions are sufficiently strong. However, these authors consider only
traps with weakly coupled wells, and therefore cannot determine the degree
of fragmentation of the ground state in the regime of low barrier strengths
where the coupling between the wells is strong. Moreover, this paper does
not address the issue of the spatial coherence of the ground state. Since
the presence of long-range order is a defining characteristic of
Bose-Einstein condensation, it is critical to understand the manners in
which this spatial coherence can be lost.

There has also been theoretical work on the problem of Bose condensates
containing atoms in two different internal states, which is analogous to the
two well problem. Esry {\it et al. }\cite{Esry} determine the probability
distributions and lifetimes of two interacting condensates in different
internal states confined to the same trap, while Cirac {\it et al. }\cite
{Cirac} as well as Steel and Collett \cite{Steel} consider the ground states
of such a system when the internal atomic states can be controlled by a
Josephson-like laser coupling. In this case, the distinguishability of atoms
in the two condensates is ensured by their internal state rather than their
single-particle wavefunctions. However, the requirement of orthogonality in
the spatial degrees of freedom is indispensable in a multiple well problem
since it is precisely the shape of the single-particle wavefunctions that
determines the degree of fragmentation in the system.

\strut The remainder of the paper is organized as follows. In section II we
present our model and define some useful measures of spatial coherence of
the condensate. In section III we present approximate analytic solutions of
our model in the regime of nearly non-interacting particles, and compare
these to a numerical solution for a particular choice of the external
potential. The experimental signature of fragmentation and finite
temperature effects are discussed in section V, followed by our concluding
remarks in section VI.

\section{The Model}

\subsection{The basic approach}

Our system consists of an even number, $N,$ of spinless bosons at zero
temperature. We model the interactions by a two-particle pseudopotential in
the shape-independent approximation, $V({\bf r},{\bf r}^{^{\prime
}})=g\delta ({\bf r}-{\bf r}^{^{\prime }})$ with an interaction strength $%
g=4\pi a_{sc}\hbar ^{2}/m$, where $a_{sc}$ is the s-wave scattering length,
and $m$ is the mass of the bosons. The Hamiltonian is given by \cite{Fetter
and Walecka} 
\begin{equation}
\hat{H}=\int d^{3}{\bf r}\left[ 
\begin{array}{c}
-\frac{\hbar ^{2}}{2m}\hat{\Psi}^{\dag }({\bf r}){\bf \nabla }^{2}\hat{\Psi}(%
{\bf r})+U({\bf r})\hat{\Psi}^{\dag }({\bf r})\hat{\Psi}({\bf r}) \\ 
+\frac{g}{2}\hat{\Psi}^{\dag }({\bf r})\hat{\Psi}^{\dag }({\bf r})\hat{\Psi}(%
{\bf r})\hat{\Psi}({\bf r})
\end{array}
\right] ,  \label{H total}
\end{equation}
where $\hat{\Psi}({\bf r})$ is the quantum field operator, and $U({\bf r})$
is the external potential. The external potential is taken to exhibit a
single minimum along the $y$ and $z$ axes, and a double minimum along the $x$
axis. It is also taken to be symmetric about $x=0.$

In order to capture the phenomenon of fragmentation in our model of the
ground state, we must go beyond a mean field analysis. Specifically, we
consider arbitrary superpositions of Fock states where up to two
single-particle states are occupied. This corresponds to postulating a state
vector of the form 
\begin{equation}
\left| \psi \right\rangle =\sum_{N_{1}=0}^{N}C_{N_{1}}\left|
N_{1},N_{2}\right\rangle _{(\phi _{1},\phi _{2})},
\label{mixed config state}
\end{equation}
where 
\[
\left| N_{1},N_{2}\right\rangle _{(\phi _{1},\phi _{2})}\equiv \frac{%
(a_{1}^{\dag })^{N_{1}}}{\sqrt{N_{1}}}\frac{(a_{2}^{\dag })^{N_{2}}}{\sqrt{%
N_{2}}}\left| vac\right\rangle 
\]
is the Fock state in which $N_{1}$ particles occupy the single-particle
state $\phi _{1}({\bf r})=\left\langle {\bf r}\right| a_{1}^{\dag }\left|
vac\right\rangle $ and $N_{2}$ particles occupy $\phi _{2}({\bf r}%
)=\left\langle {\bf r}\right| a_{2}^{\dag }\left| vac\right\rangle .$ The
total number of particles is fixed, $N_{2}\equiv N-N_{1},$ the vector
consisting of the set of coefficients $C_{N_{1}}$ is normalized$,$ and the
single-particle wavefunctions $\phi _{1}$ and $\phi _{2}$ are both
normalized and orthogonal to one another.

The state (\ref{mixed config state}) is certainly not the most general state
one can consider. Indeed, a state such as (\ref{mixed config state}) would
be a poor choice if one were interested in studying the depletion of a
single condensate due to interactions, since one there expects a certain
fraction of the particles to be distributed among a macroscopic number of
single-particle states. However, in this paper we are interested in the
possibility of the particles being redistributed into a few single-particle
states that are each macroscopically occupied. We restrict ourselves to {\it %
two }single-particle states because the double-well geometry we are
considering encourages fragmentation into two pieces \cite{lanl}. Although
it may be energetically favorable to fragment into more than two pieces at
very high particle densities, we defer consideration of this possibility to
a later work.

Among the many-body states defined by (\ref{mixed config state}), we
consider only those which have the same symmetry as the Hamiltonian under
reflections about $x=0$. This implies that the single-particle wavefunctions
are mirror images of one another across $x=0$ within a phase factor, $\phi
_{1}(-x,y,z)=$ $e^{i\theta }\phi _{2}(x,y,z),$ and that the coefficients
satisfy $C_{N_{1}}=C_{N-N_{1}}.$ With this assumption, and choosing $\phi
_{1}$ and $\phi _{2}$ to be real, the Hamiltonian takes the form 
\begin{eqnarray}
\hat{H}_{2} &=&\epsilon _{11}\hat{N}+\left( \epsilon _{12}+gT_{1}(\hat{N}%
-1)\right) \left( a_{1}^{\dag }a_{2}+a_{2}^{\dag }a_{1}\right)  \nonumber \\
&&+\frac{gT_{0}}{2}\left( \hat{N}_{1}^{2}+\hat{N}_{2}^{2}-\hat{N}\right) 
\nonumber \\
&&+\frac{gT_{2}}{2}(a_{1}^{\dag }a_{1}^{\dag }a_{2}a_{2}+a_{2}^{\dag
}a_{2}^{\dag }a_{1}a_{1}+4\hat{N}_{1}\hat{N}_{2}),  \label{H two wf}
\end{eqnarray}
where $\hat{N}_{1}=a_{1}^{\dag }a_{1},\hat{N}_{2}=a_{2}^{\dag }a_{2},$ $\hat{%
N}=\hat{N}_{1}+\hat{N}_{2},$ and where 
\begin{eqnarray*}
\epsilon _{11} &=&\int d^{3}r\phi _{1}({\bf r})\left( -\frac{\hbar ^{2}}{2m}%
{\bf \nabla }^{2}+U({\bf r})\right) \phi _{1}({\bf r}), \\
\epsilon _{12} &=&\int d^{3}r\phi _{1}({\bf r})\left( -\frac{\hbar ^{2}}{2m}%
{\bf \nabla }^{2}+U({\bf r})\right) \phi _{2}({\bf r}), \\
T_{0} &=&\int d^{3}r\phi _{1}^{4}({\bf r}), \\
T_{1} &=&\int d^{3}r\phi _{1}^{3}({\bf r})\phi _{2}({\bf r}), \\
T_{2} &=&\int d^{3}r\phi _{1}^{2}({\bf r})\phi _{2}^{2}({\bf r}).
\end{eqnarray*}
These quantities have the following physical interpretation: $\epsilon _{11}$
is the single-particle energy for the state $\phi _{1}$; $\epsilon _{12}$ is
proportional to the inversion frequency of a single particle in the external
potential; finally, $T_{0}$ quantifies the self-interaction energy, while $%
T_{1}$ and $T_{2}$ both quantify the cross-interaction energy.

In order to facilitate comparison of our work with earlier studies \cite
{Milburn},\cite{Steel}, we re-express the Hamiltonian in terms of operators
satisfying angular momentum commutation relations, rather than in terms of
the creation and annihilation operators we have employed thus far. We
introduce the operators 
\begin{eqnarray*}
\hat{J}_{z} &=&\frac{1}{2}(a_{2}^{\dag }a_{1}+a_{1}^{\dag }a_{2}), \\
\hat{J}_{y} &=&\frac{i}{2}(a_{2}^{\dag }a_{1}-a_{1}^{\dag }a_{2}), \\
\hat{J}_{x} &=&\frac{1}{2}(\hat{N}_{2}-\hat{N}_{1}),
\end{eqnarray*}
which form an angular momentum algebra with total angular momentum $j=N/2$ 
\cite{Milburn}. In terms of these operators, the Hamiltonian can be
rewritten as 
\begin{equation}
\begin{array}{c}
\hat{H}_{2}=E_{0}+2\left( \epsilon _{12}+gT_{1}(N-1)\right) \hat{J}%
_{z}+2gT_{2}\hat{J}_{z}^{2} \\ 
+g(T_{0}-T_{2})\hat{J}_{x}^{2},
\end{array}
\label{H angmom}
\end{equation}
where 
\[
E_{0}\equiv \epsilon _{11}N+\frac{1}{8}N(N+2)(gT_{0}+gT_{2}), 
\]
and where we have used $\hat{J}_{x}^{2}+\hat{J}_{y}^{2}+\hat{J}%
_{z}^{2}=j(j+1)$ to eliminate $\hat{J}_{y}^{2}$ from the expression. The
observable corresponding to $\hat{J}_{x}$ is the particle number difference
between the localized states $\phi _{1}$ and $\phi _{2}$. Defining the
wavefunctions $\phi _{s}=2^{-1/2}(\phi _{1}+\phi _{2})$ and $\phi
_{a}=2^{-1/2}(\phi _{1}-\phi _{2}),$ which are respectively symmetric and
antisymmetric about $x=0,$ one sees that $\hat{J}_{z}$ can be rewritten as $%
\frac{1}{2}(a_{s}^{\dag }a_{s}-a_{a}^{\dag }a_{a}),$ where $a_{s}^{\dag }$
and $a_{a}^{\dag }$ are the creation operators associated with $\phi _{s}$
and $\phi _{a}$ respectively. Thus, $\hat{J}_{z}$ corresponds to the
particle number difference between the symmetrized states $\phi _{s}$ and $%
\phi _{a}.$ Finally, $\hat{J}_{y}$ corresponds to the condensate momentum.
Since we are considering the ground state it follows that $\left\langle
J_{y}\right\rangle =0,$ and since the ground state is symmetric under
reflections about $x=0$ it follows that $\left\langle J_{x}\right\rangle =0.$

\subsection{A restricted variational principle}

We now turn to the problem of identifying the ground state of our model.
This can be achieved by minimizing the expectation value of $\hat{H}_{2}$
with respect to variations in both the coefficients $C_{N_{1}}$ and the
single-particle wavefunction $\phi _{1},$ subject to the constraints that
the set of coefficients is normalized, and $\phi _{1}$ is both normalized
and orthogonal to $\phi _{2},$ its mirror image about $x=0.$ However, it is
in fact more convenient to minimize the expectation value of $\hat{H}_{2}$
with respect to variations in $\phi _{s}$ and $\phi _{a}$ rather than $\phi
_{1}.$ The reason for this is that no constraint corresponding to
orthogonality is required when working with $\phi _{s}$ and $\phi _{a},$
since they are orthogonal by construction; as a result the analysis is
simplified.

We begin with the variation of $\phi _{s}$ and $\phi _{a},$ implementing the
normalization constraints through Lagrange multipliers $E_{s}$ and $E_{a}$
respectively. This results in two coupled non-linear Schr\"{o}dinger
equations for $\phi _{s}$ and $\phi _{a}$

\begin{equation}
\left[ 
\begin{array}{c}
-\frac{\hbar ^{2}{\bf \nabla }^{2}}{2m}+U({\bf r})+g\Gamma _{\alpha }^{\circ
}\phi _{\alpha }^{2}({\bf r}) \\ 
+g\Gamma _{\alpha }^{x}\phi _{\beta }^{2}({\bf r})
\end{array}
\right] \phi _{\alpha }({\bf r})=E_{\alpha }\phi _{\alpha }({\bf r})\text{,}
\label{wf ev eqn}
\end{equation}
where 
\begin{eqnarray}
\Gamma _{\alpha }^{\circ } &=&\left\langle (a_{\alpha }^{\dag }a_{\alpha
})^{2}-a_{\alpha }^{\dag }a_{\alpha }\right\rangle /\left\langle a_{\alpha
}^{\dag }a_{\alpha }\right\rangle ,\text{and}  \label{gammas} \\
\Gamma _{\alpha }^{x} &=&\left\langle a_{\alpha }^{\dag }a_{\alpha }^{\dag
}a_{\beta }a_{\beta }+a_{\beta }^{\dag }a_{\beta }^{\dag }a_{\alpha
}a_{\alpha }+4a_{\alpha }^{\dag }a_{\alpha }a_{\beta }^{\dag }a_{\beta
}\right\rangle /\left\langle a_{\alpha }^{\dag }a_{\alpha }\right\rangle , 
\nonumber
\end{eqnarray}
and where the indices $(\alpha ,\beta )$ take the values $(s,a)$ and $(a,s)$.

We now minimize the expectation value of $\hat{H}_{2}$ with respect to
variations in the $C_{N_{1}},$ and implement the normalization constraint on
the $C_{N_{1}}$ through a Lagrange multiplier $E.$\ This results in a
five-term recurrence relation for the coefficients 
\begin{eqnarray}
&&\left[ N\epsilon _{11}+\frac{gT_{0}}{2}\left( N_{1}^{2}+N_{2}^{2}-N\right)
+2gT_{2}N_{1}N_{2}-E\right] C_{N_{1}}  \nonumber \\
&&+\left[ \epsilon _{12}+gT_{1}(N-1)\right] \left[ 
\begin{array}{c}
\sqrt{N_{1}(N_{2}+1)}C_{N_{1}-1} \\ 
+\sqrt{N_{2}(N_{1}+1)}C_{N_{1}+1}
\end{array}
\right]  \nonumber \\
&&+\frac{gT_{2}}{2}\left[ 
\begin{array}{c}
\sqrt{(N_{1}-1)N_{1}(N_{2}+1)(N_{2}+2)}C_{N_{1}-2} \\ 
+\sqrt{(N_{2}-1)N_{2}(N_{1}+1)(N_{1}+2)}C_{N_{1}+2}
\end{array}
\right]  \label{coef ev eqn} \\
&=&0\text{,}  \nonumber
\end{eqnarray}
for each value of $N_{1};$ $E$ is immediately identified as the expectation
value of $\hat{H}_{2}.$ \ The latter set of equations forms a matrix
eigenvalue equation for the $N$-element vector of coefficients $C_{N_{1}}.$
Given values for $\epsilon _{11},\epsilon _{12},T_{0},T_{1}$ and $T_{2},$ we
can solve this equation by diagonalizing an $N\times N$ matrix with non-zero
entries along five diagonals, a problem which is numerically tractable if
the number of non-zero coefficients is not too large.

Since equations (\ref{wf ev eqn}) and (\ref{coef ev eqn}) form a coupled set
of equations for $\phi _{s},\phi _{a}$ and the $C_{N_{1}}$, we must in
general solve these self-consistently. Of the many solutions thus obtained,
the ground state is the one which minimizes the value of $E$. However, it is
not obvious that the solution that minimizes $E_{s}$ and $E_{a}$ also
minimizes $E;$ thus it may be necessary to compare the energies of many
solutions in order to find the ground state.

\subsection{The regime of nearly non-interacting particles}

The full problem outlined above is rather complex. In this paper, we
consider only perturbative solutions of Eq. (\ref{wf ev eqn}) in the nearly
non-interacting regime, which we here define as the regime where the
interaction energy is small compared to the difference between $\epsilon
_{s}^{(1)},$ the energy of the first symmetric excited state of the external
potential, and $\epsilon _{s},$the energy of the ground state. This is
ensured by the criterion 
\begin{equation}
gNT_{0}\ll \epsilon _{s}^{(1)}-\epsilon _{s}.  \label{pert cond}
\end{equation}
Since $T_{0}$ is on the order of the inverse of the volume of the trap, this
criterion places an upper limit on the density of the condensate.

In this regime, we can treat the non-linear terms in (\ref{wf ev eqn})
perturbatively. To obtain the expectation value of $\hat{H}_{2}$ to first
order in the perturbation, we need only solve for the eigenfunctions of (\ref
{wf ev eqn}) to zeroth order. Thus, we need only solve the two linear
Schr\"{o}dinger equations

\begin{equation}
\left[ -\frac{\hbar ^{2}{\bf \nabla }^{2}}{2m}+U({\bf r})-\epsilon _{\alpha
}\right] \phi _{\alpha }({\bf r})=0\text{,}  \label{wf ev eqn linear}
\end{equation}
for $\alpha =s,$ $a$. In this case, the wavefunctions $\phi _{s}$ and $\phi
_{a}$ are simply the two lowest single-particle energy eigenfunctions of the
external potential, and the assumption of a state of the form of (\ref{mixed
config state}) corresponds to a two mode approximation.

The solutions of (\ref{wf ev eqn linear}) determine the magnitudes of $%
\epsilon _{11},\epsilon _{12},T_{0},T_{1}$ and $T_{2}$, and these
subsequently define the form of the recurrence relation (\ref{coef ev eqn})
that must be solved to obtain the coefficients $C_{N_{1}}.$ Although the
full results are presented in section III, it is illustrative to consider
the ground state of $\hat{H}_{2}$ in two particularly simple limits: that of
no barrier and that of an infinitely strong barrier.

In the absence of any barrier, $\left| \epsilon _{12}\right| \simeq \epsilon
_{s}^{(1)}-\epsilon _{s},$ and since $\left| T_{1}\right| $ and $T_{2}$ are
on the order of $T_{0}$ or less, it follows from the criterion (\ref{pert
cond}) that $\left| \epsilon _{12}\right| \gg NgT_{0},Ng\left| T_{1}\right|
,NgT_{2}$. If we provisionally assume that the ground state fulfils the
conditions that $N\left| \left\langle \hat{J}_{z}\right\rangle \right|
\gtrsim \left\langle \hat{J}_{z}^{2}\right\rangle ,\left\langle \hat{J}%
_{x}^{2}\right\rangle ,$ then we are led to approximate the Hamiltonian by 
\begin{equation}
\hat{H}_{2}\simeq E_{0}+2\epsilon _{12}\hat{J}_{z}.  \label{weak nobar H}
\end{equation}
The ground state of (\ref{weak nobar H}) is simply the Fock state $\left|
N\right\rangle _{\phi _{s}}$ which describes $N$ particles occupying the
single-particle ground state $\phi _{s}.$ Since this solution satisfies our
provisional assumption, the approximation is consistent$.$ Such a Fock state
is of course what one would expect for the ground state of a single well in
the limit of nearly non-interacting particles. When this state is written in
the form of (\ref{mixed config state}), that is, in the basis of $\left|
N_{1},N_{2}\right\rangle _{(\phi _{1},\phi _{2})}$ states, rather than the
basis of $\left| N_{s},N_{a}\right\rangle _{(\phi _{s},\phi _{a})}$ states,
the coefficients $C_{N_{1}}$ form a binomial distribution over $N_{1},$
centered at $N/2$. It seems appropriate to refer to any state of the form $%
\left| N\right\rangle _{\phi _{0}}$ for macroscopic $N$ and arbitrary $\phi
_{0},$ as a `single condensate'. In this paper, we are concerned only with
single condensates wherein the single-particle wavefunction $\phi _{0}$ is
symmetric about $x=0.$ We do not introduce any additional terminology to
distinguish such a state from one with arbitrary $\phi _{0},$ since no
confusion is likely to arise.

In the limit of infinite barrier strength, the amplitudes of $\phi _{s}$ and 
$\phi _{a}$ at $x=0$ are necessarily zero, while $\phi _{s}$ and $\phi _{a}$
at $x\ne 0$ satisfy the same equation. Consequently, $\phi _{s}$ and $\phi
_{a}$ differ only in their symmetry under reflection about $x=0,$ and $%
\epsilon _{12}=T_{1}=T_{2}=0.$ The Hamiltonian of (\ref{H angmom}) then
reduces to 
\begin{equation}
\hat{H}_{2}=E_{0}+gT_{0}\hat{J}_{x}^{2}.  \label{weak infbar H}
\end{equation}
The ground state is $\left| N/2,N/2\right\rangle _{(\phi _{1},\phi _{2})},$
which describes two independent condensates, or in other words, a condensate
which is fragmented into two pieces. Since we are considering a potential
well that is symmetric about $x=0,$ the two fragments are equally populated.
It seems appropriate to refer to any state of the form $\left|
N_{1},N_{2}\right\rangle _{(\phi _{1},\phi _{2})}$ where $N_{1}$ and $N_{2}$
are macroscopic and $\phi _{1}$ and $\phi _{2}$ are orthogonal, as a `dual
condensate' \cite{footnote1}. In this paper we will be concerned only with
dual condensates wherein $N_{1}=N_{2}=N/2,$ and $\phi _{1}$ and $\phi _{2}$
are mirror images of one another across $x=0.$

The analysis above confirms, for the limit of nearly non-interacting
particlws, the results of an earlier study \cite{lanl}: the ground state is
well approximated by a single condensate at zero barrier strength, and a
dual condensate at infinite barrier strength. At intermediate barrier
strengths, we keep all the terms in $\hat{H}_{2}$ for our calculations.
Although the cross-interaction terms are typically found to be small for
generic shapes of the double-well potential, it is not obvious that these
terms are negligible for an arbitrary potential, and thus we include them in
our analytic results wherever possible.

\subsection{Measures of the degree of fragmentation}

Finally, in order to facilitate the interpretation of our results we
highlight some observables that reveal the degree of spatial fragmentation
of the condensate. The most useful observables for this purpose are those
that probe the spatial coherence of the condensate across the barrier. In
analogy to measures of optical coherence \cite{Loudon}, we normalize the
first-order correlation function, $\rho _{1}({\bf r},{\bf r}^{\prime
})=\left\langle \hat{\Psi}^{\dag }({\bf r})\hat{\Psi}({\bf r}^{\prime
})\right\rangle $, to obtain the degree of first-order spatial coherence
between points ${\bf r}$ and ${\bf r}^{\prime },$%
\begin{equation}
g^{(1)}({\bf r},{\bf r}^{\prime })=\frac{\rho _{1}({\bf r},{\bf r}^{\prime })%
}{\left[ \rho _{1}({\bf r},{\bf r})\rho _{1}({\bf r}^{\prime },{\bf r}%
^{\prime })\right] ^{\frac{1}{2}}}.
\end{equation}
Considering points ${\bf r}=(x,y,z)$ and ${\bf r}^{\prime }=(-x,y,z)$ where $%
x$ is positive and chosen to be sufficiently large so that $\left| \phi _{1}(%
{\bf r})\right| \ll \left| \phi _{2}({\bf r})\right| $ and $\left| \phi _{1}(%
{\bf r}^{\prime })\right| \gg \left| \phi _{2}({\bf r}^{\prime })\right| ,$
for any state of the form of \ref{mixed config state} that is symmetric
under reflection about $x=0,$ the quantity $g^{(1)}({\bf r},{\bf r}^{\prime
})$ is in fact independent of ${\bf r}$ and ${\bf r}^{\prime }$, and has the
value 
\begin{equation}
{\cal C}^{(1)}=\frac{\left\langle a_{1}^{\dag }a_{2}+a_{2}^{\dag
}a_{1}\right\rangle }{N}.  \label{first order Dc}
\end{equation}
We refer to ${\cal C}^{(1)}$ simply as the degree of first-order spatial
coherence across the barrier. It is straightforward to verify that it
attains its maximum value of $1$ for a single condensate and a value of $0$
for a dual condensate.

The second-order correlation function $\rho _{2}({\bf r},{\bf r}^{\prime
})=\left\langle \hat{\Psi}^{\dag }({\bf r})\hat{\Psi}^{\dag }({\bf r}%
^{\prime })\hat{\Psi}({\bf r}^{\prime })\hat{\Psi}({\bf r})\right\rangle ,$
which is simply the normally ordered density-density correlation, can be
normalized to obtain the degree of second-order spatial coherence between
points ${\bf r}$ and ${\bf r}^{\prime },$%
\begin{equation}
g^{(2)}({\bf r},{\bf r}^{\prime })=\frac{\rho _{2}({\bf r},{\bf r}^{\prime })%
}{\left[ \rho _{2}({\bf r},{\bf r})\rho _{2}({\bf r}^{\prime },{\bf r}%
^{\prime })\right] ^{\frac{1}{2}}}.
\end{equation}
Defining ${\cal C}^{(2)}$ in a manner completely analogous to ${\cal C}%
^{(1)},$ we find 
\begin{equation}
{\cal C}^{(2)}=\frac{1-4(\frac{\Delta N_{1}}{N})^{2}}{\frac{N-2}{N}+4(\frac{%
\Delta N_{1}}{N})^{2}},  \label{second order Dc}
\end{equation}
where $\Delta N_{1}\equiv (\left\langle \hat{N}_{1}^{2}\right\rangle
-\left\langle \hat{N}_{1}\right\rangle ^{2})^{1/2}$ is the variance in the
number of particles occupying the localized state $\phi _{1}$. We refer to $%
{\cal C}^{(2)}$ as the degree of second-order spatial coherence across the
barrier. The variance in $N_{1}$ for a single condensate $\left|
N\right\rangle _{\phi _{s}}$ is that of a binomial distribution over $N_{1},$
namely $\sqrt{N}/2.$ For a dual condensate the number of particles in a well
is fixed, so that $\Delta N_{1}=0.$ As a consequence, ${\cal C}^{(2)}-1=0$
for a single condensate $\left| N\right\rangle _{\phi _{s}}$, and ${\cal C}%
^{(2)}-1=2/N$ for a dual condensate. Since $\Delta N_{1}$ is sufficient to
specify ${\cal C}^{(2)},$while being simpler to interpret, we use $\Delta
N_{1}$ together with ${\cal C}^{(1)}$ to characterize our results.

\section{Analytic Approximations and Numerical Solutions}

For a given shape of the double-well potential, it is straightforward to
obtain the single particle ground state and first excited state by solving
the linear Schr\"{o}dinger equation (\ref{wf ev eqn linear}). The localized
single-particle states $\phi _{1}$ and $\phi _{2}$ are simply the sum and
difference of the single-particle ground and first excited states. Using
these wavefunctions, the coefficients $C_{N_{1}}$ can be obtained by solving
the recurrence relation (\ref{coef ev eqn}).

\strut We here present approximate analytic solutions for the $C_{N_{1}}$
given an arbitrary shape of the double-well potential. Subsequently, we
consider a particular form of the external potential for which the single
particle ground and first excited states are obtained numerically. This
allows us to obtain a numerical solution for the $C_{N_{1}}$ and to compare
this solution to the analytic approximations.

\subsection{Continuum approximation}

\strut Suppose the coefficients for the ground state satisfy the condition 
\[
\left| C_{N_{1}+1}-C_{N_{1}}\right| \ll C_{N_{1}}. 
\]
It is then useful to construct a function $C(u),$ defined over the real
numbers, such that $C(u)=C_{N_{1}}$ at the discrete points $u=$ $%
N^{-1}(N_{1}-N/2),$ and such that $C^{\prime }(u)\ll NC(u).$ Given this
assumption of smoothness, a coefficient of the form $C_{N_{1}+p},$ where $p$
is a small integer, is well approximated by a Taylor expansion of $C(u+p/N)$
to second order in $p/N$ . In this way, the recurrence relation (\ref{coef
ev eqn}) for the $C_{N_{1}}$ can be recast as a second order differential
equation for the function $C(u).$ Moreover, any sum over $N_{1}$ can be
approximated by an integral over $u.$ In particular, the constraint of
normalization for the coefficients is replaced by the constraint that the
integral of $C^{2}(u)$ over all $u$ is $1/N.$ The assumption of smoothness
is readily verified to be appropriate for a single condensate, and we
therefore expect it to continue to hold for solutions over a range of small
barrier heights.

We also make use of the fact that the coefficients $C_{N_{1}}$ are
significant only in the region where $N_{1}\lesssim \sqrt{N},$ or
equivalently, that the function $C(u)$ is significant only where $u\lesssim 
\sqrt{1/N}.$ This follows from the fact that any set of coefficients that
has significant amplitude outside the range $N_{1}\lesssim \sqrt{N}$ also
has an energy that is larger than a single condensate; the interaction
energy is greater since it scales with $\Delta N_{1}$, and the
single-particle energy is greater since it is a minimum for a single
condensate. It is therefore appropriate to expand each of the $N_{1}$%
-dependent terms as a power series in $u:$%
\begin{eqnarray*}
\frac{2}{N}\sqrt{N_{1}(N_{2}+1)} &=&\sum_{n=0}^{\infty }I_{n}u^{n}, \\
\frac{4}{N^{2}}\sqrt{(N_{1}-1)N_{1}(N_{2}+1)(N_{2}+2)} &=&\sum_{n=0}^{\infty
}J_{n}u^{n},
\end{eqnarray*}
which implicitly defines the $I_{n}$ and $J_{n}.$

The second-order differential equation for $C(u)$ we obtain is found to have
a first-order term which can be eliminated by the substitution $\bar{C}%
(u)=C(u)\exp (-\sum_{n}a_{2n}u^{2n})$ with an appropriate choice of the
constants $a_{2n}.$ It then follows that $\bar{C}(u)$ satisfies a
second-order differential equation identical to that of a particle with
position $u$ in a one-dimensional potential well of even powers of $u.$
Since the solution is only significant in the range $u\lesssim \sqrt{1/N},$
we make the approximation that terms in the potential that are quartic or of
higher order in $u$ are small perturbations upon the quadratic term, and can
be neglected. In this case, the function $\bar{C}(u)=C(u)\exp (-\tau
u^{2}/2) $ satisfies the equation for the modes of a simple harmonic
oscillator: 
\begin{equation}
-\bar{C}^{\prime \prime }(u)+(\nu ^{2}+\tau ^{2})u^{2}\bar{C}(u)=(\eta +\tau
)\bar{C}(u),  \label{diff eqn at small u}
\end{equation}
where, after expanding the $I_{n}$ and $J_{n}$ to leading order in $1/N,$ 
\begin{eqnarray}
\eta &=&2N^{2}\frac{E/N-\left[ \epsilon _{11}+\epsilon _{12}+Ng(\frac{1}{4}%
T_{0}+T_{1}+\frac{3}{4}T_{2})\right] }{-\epsilon _{12}-Ng(T_{1}+T_{2})}, 
\nonumber \\
\nu ^{2} &=&\frac{2N^{2}}{-\epsilon _{12}-Ng(T_{1}+T_{2})}  \nonumber \\
&&\times \left[ 
\begin{array}{c}
2\left( 1-%
{\textstyle {\eta  \over N^{2}}}%
\right) \left( -\epsilon _{12}-NgT_{1}\right) \\ 
+Ng\left( T_{0}+\left( 4%
{\textstyle {\eta  \over N^{2}}}%
-3\right) T_{2}\right)
\end{array}
\right] ,\text{ and}  \nonumber \\
\tau &=&2.
\end{eqnarray}
Thus, the solution for $C(u)$ that minimizes $\eta ,$ thereby minimizing the
energy $E,$ is a Gaussian 
\begin{equation}
C(u)=\frac{1}{\sqrt{N}}\frac{1}{(2\pi )^{1/4}\sqrt{\sigma }}e^{-\frac{u^{2}}{%
4\sigma ^{2}}},
\end{equation}
with width 
\begin{equation}
\sigma =\frac{1}{2\sqrt{N}}\frac{1}{\sqrt{-A+\sqrt{A^{2}+B}}},  \label{sigma}
\end{equation}
where 
\begin{eqnarray*}
A &=&\frac{3}{2N}\frac{-\epsilon _{12}-Ng(T_{1}+\frac{4}{3}T_{2})}{-\epsilon
_{12}-Ng(T_{1}+T_{2})},\text{and} \\
B &=&\frac{-\epsilon _{12}-Ng(T_{1}+\frac{3}{2}T_{2}-\frac{1}{2}T_{0})}{%
-\epsilon _{12}-Ng(T_{1}+T_{2})},
\end{eqnarray*}
and with energy 
\begin{equation}
E=N\left[ 
\begin{array}{c}
\frac{-\epsilon _{12}-Ng(T_{1}+T_{2})}{4N^{2}\sigma ^{2}}+(\epsilon
_{11}+\epsilon _{12}) \\ 
+Ng(\frac{1}{4}T_{0}+T_{1}+\frac{3}{4}T_{2})
\end{array}
\right] .
\end{equation}
The variance in $N_{1}$ for this solution is simply $N\sigma ,$ while the
degree of first-order spatial coherence is given by 
\begin{equation}
{\cal C}^{(1)}=e^{-\frac{1}{8\sigma ^{2}N^{2}}}(1+%
{\textstyle {1 \over N}}%
-2\sigma ^{2}).  \label{cntm Dc}
\end{equation}

\strut In the absence of any barrier, $\left| \epsilon _{12}\right| \gg
NgT_{0},Ng\left| T_{1}\right| ,NgT_{2},$ and it can be verified that eq. (%
\ref{sigma}) predicts the appropriate value for this limit, namely $\Delta
N_{1}\simeq 1/2\sqrt{N},$ the value for a single condensate. As the barrier
strength is increased, the magnitudes of $\epsilon _{12},T_{1}$ and $T_{2}$
decrease while the magnitude of $T_{0}$ does not vary significantly; it
therefore follows from (\ref{sigma}) that $\Delta N_{1}$ will decrease with
barrier strength. \strut When $\Delta N_{1}$ falls below $1,$ the assumption
of smoothness breaks down. Thus the range of validity of the continuum
approximation is $\Delta N_{1}\gtrsim 1,$ or equivalently, ${\cal C}%
^{(1)}\gtrsim 0.88.$

If the potential is such that $Ng\left| T_{1}\right| ,NgT_{2}\ll \left|
\epsilon _{12}\right| $ continues to hold as the barrier strength is raised
from zero, then to leading order in $1/N$ the expression for $\sigma $
simplifies to 
\[
\sigma =\frac{1}{2\sqrt{N}}\frac{1}{\sqrt{1+\frac{N}{2}\frac{gT_{0}}{%
(-\epsilon _{12})}}}, 
\]
depending only on the ratio of the interaction energy to the splitting
between the symmetric and antisymmetric levels of the trap.

\subsection{Two-coefficient approximation}

At large barrier strengths, we make use of the following conditions: 
\begin{eqnarray}
\left| \gamma \right| &\ll &1,\text{ where}  \nonumber \\
\gamma &\equiv & 
{\displaystyle {\sqrt{\frac{N}{2}\left( \frac{N}{2}+1\right) }(-\epsilon _{12}-g(N-1)T_{1}) \over gT_{0}}}%
,  \label{gamma}
\end{eqnarray}
and 
\begin{eqnarray}
\left| \zeta \right| &\ll &1,\text{ where}  \nonumber \\
\zeta &\equiv & 
{\displaystyle {N^{2}gT_{2} \over \sqrt{\frac{N}{2}\left( \frac{N}{2}+1\right) }(-\epsilon _{12}-g(N-1)T_{1})}}%
.  \label{zeta}
\end{eqnarray}
The first of these conditions is always satisfied for sufficiently strong
barriers since in the limit of an infinitely strong barrier, $\epsilon
_{12}=T_{1}=0,$ while $T_{0}$ is finite. Moreover, we have numerically
verified that the second condition holds at sufficiently large barrier
strengths for a variety of external potentials. Within the domain of
applicability of these conditions, we seek coefficients $C_{N_{1}}$ that
satisfy the recurrence relation (\ref{coef ev eqn}) to first order in $%
\gamma $ and $\zeta .$ Dividing the recurrence relation by $gT_{0},$ one
finds that all terms involving $N^{2}gT_{2}$ have a magnitude on the order
of $\gamma \zeta $ and can therefore be neglected. In this limit, the
following set of coefficients are a solution:

\begin{equation}
\begin{array}{lll}
C_{N_{1}} & =\sqrt{1-2\gamma ^{2}} & \text{for }N_{1}=\frac{N}{2} \\ 
& =\gamma & \text{for }N_{1}=\frac{N}{2}+1,\frac{N}{2}-1 \\ 
& =0\text{ } & \text{otherwise.}
\end{array}
\label{2coef state}
\end{equation}
The energy in this case is 
\begin{equation}
E=N\epsilon _{11}+\left( \frac{N(N-2)}{4}-2\gamma ^{2}\right) gT_{0}.
\label{2coef energy}
\end{equation}
We refer to this approximation as the {\it two-coefficient approximation,}
and we dub any state of the form of (\ref{2coef state}) a {\it perturbed
dual condensate}. For such a state, ${\cal C}^{(1)}$ and $\Delta N_{1}$ are
given by 
\begin{eqnarray}
{\cal C}^{(1)} &=&2\gamma \sqrt{1-2\gamma ^{2}}\sqrt{1+%
{\textstyle {2 \over N}}%
},\text{ and}  \label{2coef Dc} \\
\Delta N_{1} &=&\sqrt{2}\gamma .  \label{2coef deltaN1}
\end{eqnarray}
Keeping terms to first order in $\gamma $, and to leading order in powers of 
$1/N,$ we have ${\cal C}^{(1)}=2\gamma .$ The range of validity of the
two-coefficient approximation is the range of barrier strengths for which $%
{\cal C}^{(1)}\ll 1$ and $\Delta N_{1}\ll 1.$

An alternative manner of deriving this solution that is perhaps more
physically intuitive, is to begin by assuming a state of the form of (\ref
{2coef state}) and showing that the value of $\gamma $ that minimizes the
energy is indeed the value given in (\ref{gamma}). We begin by recalling the
form of $\hat{H}_{2},$ exhibited in (\ref{H two wf}). Assuming that $\left|
\zeta \right| \ll 1$, the $T_{2}$ term in $\hat{H}_{2}$ can safely be
neglected. If we introduce the operators 
\begin{eqnarray*}
\hat{n}_{1} &=&\hat{N}_{1}-\frac{N}{2},\text{ and} \\
\hat{n}_{2} &=&\hat{N}_{2}-\frac{N}{2},
\end{eqnarray*}
then we find that 
\begin{eqnarray*}
\left\langle \hat{H}_{2}\right\rangle &=&E_{dual}+(\epsilon
_{12}+gT_{1}(N-1))\left\langle a_{1}^{\dag }a_{2}+a_{2}^{\dag
}a_{1}\right\rangle \\
&&+\frac{1}{2}gT_{0}(\left\langle \hat{n}_{1}^{2}\right\rangle +\left\langle 
\hat{n}_{2}^{2}\right\rangle ),
\end{eqnarray*}
where $E_{dual}$ is the energy of the dual condensate. Since the magnitude
of the cross-interaction term involving $T_{1}$ only depends on the
many-body state through expectation values of bilinear operators, this term
together with the $\epsilon _{12}$ term can be considered as an effective
single-particle energy. To first order in $\gamma $ and to leading order in $%
1/N,$ the perturbed dual condensate has $\left\langle a_{1}^{\dag
}a_{2}+a_{2}^{\dag }a_{1}\right\rangle =2\gamma N$ and consequently it has
an effective single-particle energy benefit over the dual condensate of $%
2\gamma N(-\epsilon _{12}-gNT_{1})$ (this quantity is positive at the
barrier strengths of interest, since $\epsilon _{12}<0$ and typically $%
T_{1}<0)$. On the other hand, $\left\langle \hat{n}_{1}^{2}\right\rangle
=\left\langle \hat{n}_{2}^{2}\right\rangle =2\gamma ^{2}$ for such a state,
corresponding to a self-interaction energy cost of $2\gamma ^{2}gT_{0}$.
Thus, the effective single-particle energy of the perturbed dual condensate
decreases linearly with $\gamma ,$ while the self-interaction energy
increases quadratically with this parameter. The minimum occurs precisely
when $\gamma $ has the value $N(-\epsilon _{12}-gNT_{1})/2gT_{0},$ which
approximates the value in (\ref{gamma}) for $N\gg 1.$

For typical double-well potentials, where $Ng\left| T_{1}\right| \ll \left|
\epsilon _{12}\right| ,$ $\gamma $ is well approximated by $N(-\epsilon
_{12})/2gT_{0}$ to leading order in $1/N.$ In this case, the transition from
a completely fragmented condensate to one that shows some coherence across
the barrier occurs when the number of particles times the ratio of the
inversion frequency to the self-interaction energy becomes non-negligible.
Thus, for a given shape of the trap and a fixed barrier strength, the degree
of fragmentation decreases as the number of particles is increased but
increases as the strength of the interaction is increased.

\subsection{Numerical solutions}

The form of the double-well potential in the MIT condensate interference
experiment \cite{Andrews et al.} is well modelled by a term that is harmonic
along all three Cartesian axes, with frequencies $\omega _{x},$ $\omega _{y}$
and $\omega _{z}$ respectively, to which is added a Gaussian barrier of
width $\delta $ and strength $\alpha $ centered at $x=0$, 
\begin{equation}
U({\bf r})=m(\omega _{x}^{2}x^{2}+\omega _{y}^{2}y^{2}+\omega _{z}^{2}z^{2})+%
\frac{\alpha }{\sqrt{2\pi }\delta }e^{-\frac{x^{2}}{2\delta ^{2}}}.
\label{external potential}
\end{equation}
The parameter values appropriate for Ref. \cite{Andrews et al.} are $\omega
_{x}=2\pi \times 19$ Hz, $\omega _{y}=\omega _{z}=2\pi \times 250$ Hz, and $%
\delta =6$ $\mu $m. Within such a trap, the criterion (\ref{pert cond}) for
the applicability of our perturbative approach becomes $N\ll 100$, which is
much smaller than the number of condensate atoms in their experiment. We
consider instead a larger trap, specifically, one which is isotropic with
the trapping frequency of the axis of weakest confinement in the MIT trap, $%
\omega _{x}=$ $\omega _{y}=\omega _{z}=2\pi \times 19$ Hz. In this case our
perturbative approach is good for up to approximately $N=100$ particles, and
this is the example we consider. The scattering length of $^{23}$Na is taken
to be $a_{sc}=3$ nm \cite{footnote RNSW}. \strut

In Fig. 1 we plot the profile along the $x$-axis of the external potential $%
U({\bf r})$ and the wavefunctions $\phi _{1}({\bf r})$ and $\phi _{2}({\bf r}%
),$ together with the coefficients $C_{N_{1}}$ for several values of the
barrier strength. Fig. 2 displays the degree of first order spatial
coherence, ${\cal C}^{(1)},$ and the variance, $\Delta N_{1},$ in the number
of particles occupying the localized state $\phi _{1}$ as a function of the
barrier strength. The generic features of these results persist for
different choices of parameters in (\ref{external potential}) as well as for
different choices of the form of $U({\bf r}).$

Also displayed in Fig. 2 are the values for ${\cal C}^{(1)}$ and $\Delta
N_{1}$ given by the continuum approximation, as specified by eqs. (\ref
{sigma}) and (\ref{cntm Dc}), and given by the two-coefficient
approximation, as specified by eqs. (\ref{2coef Dc}) and (\ref{2coef deltaN1}%
), for the same choice of external potential. \strut For their respective
ranges of validity, the analytic approximations are found to fit the
numerical results extremely well.

From these calculations arise the following picture of the transition
between a single and a fragmented condensate. Moving up from zero barrier
strength, there is a range of barrier strengths over which ${\cal C}^{(1)}$
is close to unity, while $\Delta N_{1}$ falls from its single condensate
value of $1/2\sqrt{N}$ to a value of $1.$ Moving down from infinite barrier
strength, there is a range of barrier strengths over which ${\cal C}^{(1)}$
and $\Delta N_{1}$ are both much less than $1.$ Between these two domains,
there is a narrow range of barrier strengths wherein the greatest part of
the transition in ${\cal C}^{(1)}$ is made. The barrier strengths delimiting
these domains can be estimated analytically using the approximations
presented in this section.

\section{Discussion}

\subsection{The experimental signature of fragmentation}

Herein we consider a measurement of the first-order degree of spatial
coherence. This is accomplished by a type of interference experiment that
has been widely discussed in the literature \cite{Andrews et al.}\cite
{Javanainen and Wilkens}\cite{Javanainen and Yoo}\cite{CGNZ}\cite{NWSCZ}.
Essentially, it constitutes a double-slit experiment for Bose condensates.
The thought experiment runs as follows. After preparation of the condensate,
the trap potential is removed and the atoms fall under the force of gravity
through a pair of slits, located symmetrically about $x=0.$ These slits can
be formed by changing the shape of the trapping potential, as long as this
change is not so rapid that excitations are induced, and not so slow that
the system has time to relax to a new many-body ground state. For simplicity
of the analysis, we also assume that the particles on the left and right are
each given momentum translations of magnitude $\hbar k$ towards one another 
\cite{Javanainen and Wilkens}. In the absence of such translations, the
interference pattern is simply more complicated, and has been studied by
R\"{o}rhl {\em et al.} \cite{Rohrl}. We make the approximation that the
inter-particle interactions are insignificant during this expansion period.
In this case, only the single-particle wavefunctions evolve, while the
coefficients $C_{N_{1}}$ in the many-body state (\ref{mixed config state})
remain unchanged.

Suppose the slits are centered at points ${\bf r}=(x,y,z)$ and ${\bf r}%
^{\prime }=(-x,y,z),$ where $x$ is positive and chosen to be sufficiently
large so that $\left| \phi _{1}({\bf r})\right| \ll \left| \phi _{2}({\bf r}%
)\right| $ and $\left| \phi _{1}({\bf r}^{\prime })\right| \gg \left| \phi
_{2}({\bf r}^{\prime })\right| .$ In this case, the single-particle
wavefunctions $\phi _{1}({\bf r})$ and $\phi _{2}({\bf r})$ evolve to
wavefunctions localized entirely at just one of the slits. After the
momentum translation and a period of free expansion the single-particle
wavefunctions originating from the left and right of the barrier acquire
complex phase factors of $e^{ikx}$ and $e^{-ikx}$ and magnitudes we denote
by $\tilde{\phi}_{1}({\bf r})$ and $\tilde{\phi}_{2}({\bf r})$ respectively.
For many-body states that are symmetric under a reflection about $x=0,$
these magnitudes are roughly uniform and of equal magnitude in the far field
of the double-slit, so that the many-body state in the far field can be
approximated by the many-body state prior to removal of the trap, with $\phi
_{1}({\bf r})$ and $\phi _{2}({\bf r})$ replaced by $e^{ikx}$ and $e^{-ikx}$
respectively.

We now imagine detectors in the far field that are assumed to remove atoms
from the condensate \cite{Javanainen and Yoo}. The probability distribution
over the position, ${\bf r}_{1},$ of the first detection is given by the
expectation value of normally ordered field operators $P^{1}({\bf r}_{1})=%
\frac{1}{N}\left\langle \hat{\Psi}^{\dag }({\bf r}_{1})\hat{\Psi}({\bf r}%
_{1})\right\rangle .$ The probability distribution over the positions ${\bf r%
}_{1},{\bf r}_{2},...,{\bf r}_{m}$ of the first $m$ detections is given by $%
P^{m}({\bf r}_{1},{\bf r}_{2},...,{\bf r}_{m})=\frac{(N-m)!}{N!}\left\langle 
\hat{\Psi}^{\dag }({\bf r}_{1})\hat{\Psi}^{\dag }({\bf r}_{2})...\hat{\Psi}%
^{\dag }({\bf r}_{m})\hat{\Psi}({\bf r}_{m})...\hat{\Psi}({\bf r}_{2})\hat{%
\Psi}({\bf r}_{1})\right\rangle .$ The density of detection events that
emerges in a single run of the double-slit experiment has the form $\rho
^{m}({\bf r})=\frac{N}{m}\sum_{i=1}^{m}\delta ({\bf r}-{\bf r}_{i})$ where
the set of positions $\{{\bf r}_{i}\}$ are obtained from the probability
distribution $P^{m}({\bf r}_{1},{\bf r}_{2},...,{\bf r}_{m}).$ Of course,
the finite resolution of any realistic detector can be accounted for by
replacing the delta function in this expression with a suitably broadened
distribution; as long as the resolution is finer than the distance between
the fringes of the interference pattern, the difference will not be
significant.

In a single run of the double-slit experiment, both the single and dual
condensates typically yield a distribution $\rho ^{m}({\bf r})$ with
essentially the maximum possible fringe visibility. This is obviously true
for a single condensate, and has been shown to be true for a dual condensate
in the seminal paper of Javanainen and Yoo \cite{Javanainen and Yoo}. Thus,
the mere presence of such interference is not indicative of a non-zero
degree of first-order coherence. However, suppose the experiment is repeated
many times with the same initial many-body state. In this case, the spatial
phase of the interference pattern will vary randomly from one run to the
next if the initial state is a dual condensate, while it will remain fixed
if the initial state is a single condensate \cite{Javanainen and Yoo}. We
therefore expect the degree of first-order spatial coherence to be revealed
by the {\em variance} in the spatial phase of the interference pattern over
many runs, or equivalently, the fringe visibility of the {\em average}
detection pattern over many runs. Indeed, if one averages the pattern of
detections from an infinite number of runs of the double-slit experiment,
all prepared initially in the same many-body state, and each involving $m$
detection events, one obtains: 
\begin{eqnarray*}
\bar{\rho}^{m}({\bf r}) &=&\int d^{3}r_{1}...d^{3}r_{m}(\frac{N}{m}%
\sum_{i=1}^{m}\delta ({\bf r}-{\bf r}_{i}))P^{m}({\bf r}_{1},{\bf r}_{2},...,%
{\bf r}_{m}) \\
&=&\frac{N}{m}\sum_{i=1}^{m}\int
d^{3}r_{1}...d^{3}r_{i-1}d^{3}r_{i+1}...d^{3}r_{m} \\
&&\times P^{m}({\bf r}_{1},...,{\bf r}_{i-1},{\bf r},{\bf r}_{i+1},...,{\bf r%
}_{m}) \\
&=&\left\langle \hat{\Psi}^{\dag }({\bf r})\hat{\Psi}({\bf r})\right\rangle .
\end{eqnarray*}
The final equality follows from the fact that each element of the sum is
simply equal to $P^{1}({\bf r}_{i}).$ If the fringe visibility of the
average detection pattern, $\bar{\rho}^{m}({\bf r}),$ is evaluated for the
many-body state in the far field, it is found to be precisely equal to our
definition of ${\cal C}^{(1)},$ the degree of first-order spatial coherence
across the barrier, and thus enables a measurement of the latter.

Another possibility for an experimental study of fragmentation is a
measurement of the degree of second-order spatial coherence; this may be
accessible through non-resonant imaging \cite{Meystre}.

\subsection{Finite temperatures}

We close this section with a few comments on the effect of finite
temperatures on the coherence properties of the condensate. The excited
states of the system are not in general well approximated by a state of the
form of (\ref{mixed config state}). Nonetheless it can be shown that at
infinite barrier strength the first-order degree of spatial coherence across
the barrier is zero for thermal equilibrium at any finite temperature. The
proof is as follows. For an infinitely strong barrier the Hamiltonian is
separable into two terms each involving only operators pertaining to
particles on one side of the barrier. Consequently, any non-degenerate
energy eigenstates are tensor products of states describing particles on one
side of the barrier only. For any set of energy eigenstates that are
degenerate, the subspace of Hilbert space spanned by this set has a basis of
such product states. Thus, one can always identify a basis of energy
eigenstates that are product states of this sort. Since the density operator
that represents thermal equilibrium, $\hat{\rho},$ is a mixture of these
energy eigenstates, we will necessarily have $Tr(\hat{\rho}\hat{\Psi}^{\dag
}({\bf r})\hat{\Psi}({\bf r}^{\prime }))=0$ if $r$ and $r^{\prime }$ are on
opposite sides of the barrier, and consequently the degree of first-order
spatial coherence across the barrier for such a mixed state is zero as well.
In the absence of any barrier, as long as the temperature is small enough
that most of the particles are in the lowest single-particle energy level,
the first-order degree of spatial coherence for the thermal state should be
close to unity. For such temperatures, if the barrier strength is varied
from zero to infinity, the first-order degree of spatial coherence of the
thermal state will vary from nearly unity to zero. Thus, a significant
transition must still occur at such temperatures.

\section{Conclusions}

A theoretical treatment of fragmentation in Bose condensates must go beyond
a mean-field analysis. For the case of repulsive inter-particle interactions
and a double-well trapping potential, we have proposed an approach wherein
an approximation to the many-body ground state is obtained by a restricted
variational principle. We have implemented this proposal for the case of
nearly non-interacting particles.

The coherence properties that we have considered are the degrees of
first-order and second-order coherence across the central barrier of the
potential. The first of these quantifies the variance over many runs in the
spatial phase of the fringe pattern arising from the interference of atoms
on either side of the barrier. The second is essentially the density-density
correlation across the barrier, and for the states we consider it is a
simple function of the variance in the number of particles in one of the
wells. We find that as the barrier strength is increased, this variance is
continuously squeezed down from its value for a single condensate. The
degree of first-order spatial coherence is close to unity when this variance
is greater than one, but thereafter drops off rapidly. Above a certain
critical barrier strength we find that both quantities become much less than
one, indicating that the condensate is essentially completely fragmented.

We have discussed how the degree of first-order coherence might be measured
through interference experiments, and argued that a significant effect
should be present even at finite temperatures. A concern, however, is that
the ground state might be difficult to prepare if the relaxation time of the
system is long compared to the lifetime of the condensate. This could arise
if the only way for the particles to be redistributed across the barrier is
by tunneling through it. However for numbers of particles that are not too
large, this tunneling time need not be restrictive. For instance, in the
example presented in Section III.C, the single particle tunneling time at
the barrier strength where ${\cal C}^{(1)}=0.88$ is approximately one
minute, while at the barrier strength where ${\cal C}^{(1)}=0.1$ it is
roughly one hour.

Our variational approach can be extended in a straightforward manner to the
determination of the many-body ground state in systems where the external
potential has an arbitrary number, $n,$ of minima. In such cases, one would
simply introduce states that are arbitrary superpositions of Fock states
where up to $n$ single-particle wavefunctions are occupied. Such an analysis
should be of relevance to the determination of the coherence properties of
Bose condensates in optical lattices. Moreover, at extremely high densities,
where the interaction energy is dominant, it may become energetically
favorable for a condensate to begin to fragment even in the presence of a
perfectly uniform potential. We hope to address these possibilities in
future work.

\section{Acknowledgments}

\strut We wish to thank Allan Griffin, Aephraim Steinberg, Mike Steel and
Wolfgang Ketterle for helpful discussions. This work was supported by the
National Sciences and Engineering Research Council of Canada, and Photonics
Research Ontario.

\strut

\bigskip

Fig.1. Numerical solutions for the single-particle wavefunctions $\phi _{1}(%
{\bf r})$ and $\phi _{2}({\bf r})$ and the coefficients $C_{N_{1}}$ for $%
N=100$ particles and double-well potentials with barrier strengths of $%
\alpha =0,15,30,45,$ and $60$ in units of $\sqrt{\hbar ^{3}\omega _{x}/m}$.
The remaining parameter values are specified in the text. The dotted curve
is the external potential along the $x$-axis, $U(x,y=0,z=0),$ in units of $%
\hbar \omega _{x}.$ The solid and dashed curves are $\phi _{1}(x,y=0,z=0)$
and $\phi _{2}(x,y=0,z=0)$ respectively, in arbitrary units.

Fig.2. (a) The degree of first-order spatial coherence across the barrier, $%
{\cal C}^{(1)},$ and (b) the variance in the occupation number of one of the
wells$,$ $\Delta N_{1},$ as a function of the barrier strength, $\alpha $,
in units of $\sqrt{\hbar ^{3}\omega _{x}/m}$ for $N=100$ particles and the
parameter values specified in the text. The solid curve is the numerical
solution, the dashed curve is the continuum approximation, and the dotted
curve is the two-coefficient approximation.


\begin{references}
\bibitem{footnote1}  The term `double condensate' has been used to denote
two condensates of atoms of different internal spin states \cite{Esry}. We
use the term `dual condensate' for two condensates that are only
distinguished by the spatial wavefunctions of the atoms.
\end{references}
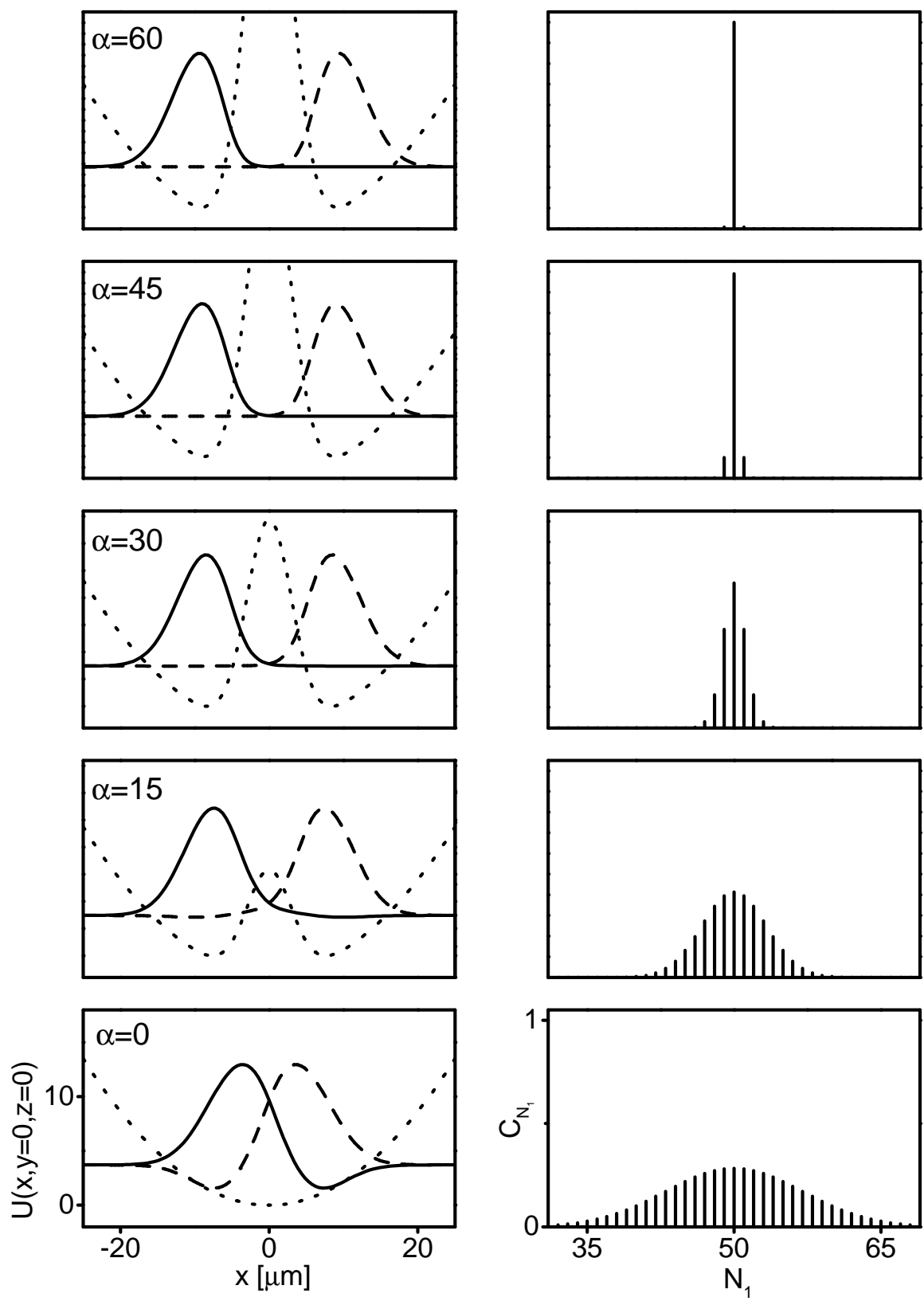
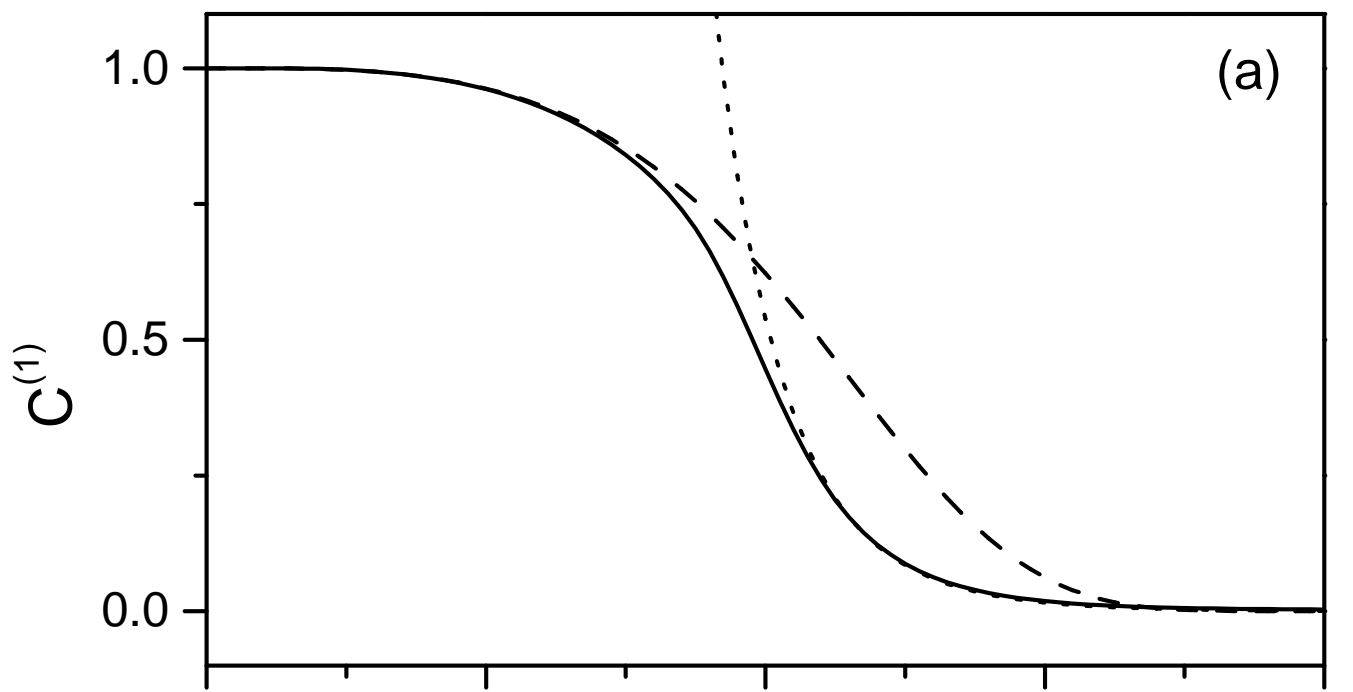
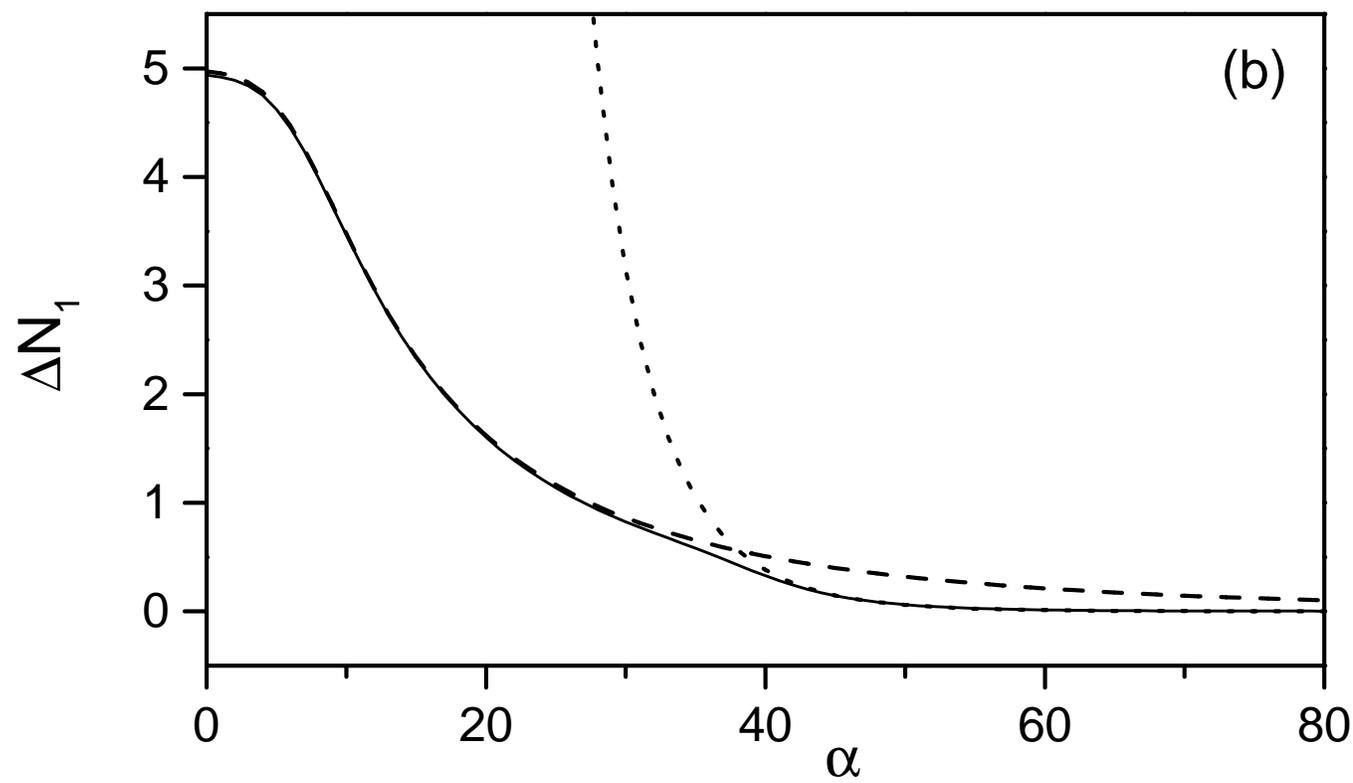
\end{document}